\documentstyle[psfig]{mn}
\def\H0{{\it H}$_0$}

\def\Ls{{\it L}$_\odot$}
\def\q0{{\it q}$_0$}
\def\kmps{km~s$^{-1}$}

\def\ergps{erg~s$^{-1}$}
\def\kmpspMpc{km~s$^{-1}$~Mpc$^{-1}$}

\def\nH{$N_{\rm H}$\thinspace} 

\def\psqcm{cm$^{-2}$}
\def\ergpspsqcm{erg~cm$^{-2}$~s$^{-1}$}

\def\cps{ct\thinspace s$^{-1}$}

\def\phpspsqcm{ph\thinspace s$^{-1}$\thinspace cm$^{-2}$}

\title[Obscured nucleus of Tol 0109--383] 
{Nuclear obscuration in the high-ionization Seyfert 2 galaxy Tol 0109--383} 
\author[K. Iwasawa et al] 
{\parbox[]{6.5in} {K.~Iwasawa$^1$, G. Matt$^2$, A.C.~Fabian$^1$, S. Bianchi$^2$, W.N.~Brandt$^3$, M.~Guainazzi$^4$, T.~Murayama$^5$ and Y. Taniguchi$^{5,6}$}\\
\\
$^1$Institute of Astronomy, Madingley Road, Cambridge CB3 0HA\\ 
$^2$Dipartimento di Fisica, Universita' degli Studi Roma Tre, Via della Vasca Navale 84, I-00146 Roma, Italy\\
$^3$Department of Astronomy \& Astrophysics, The Pennsylvania State University, 525 Davey Lab, University Park, PA16802, USA\\
$^4$XMM-Newton SOC, VILSPA, ESA, Apartado 50727, E-28080 Madrid, Spain\\
$^5$Astronomical Institute, Tohoku University, Aoba, Sendai 980-8678, Japan\\
$^6$Institute for Astronomy, University of Hawaii, 2680 Woodlawn Drive, Honolulu, HI 96822, USA
}
\date{}

\begin{document}

\maketitle

\begin{abstract}
We report the BeppoSAX detection of a hard X-ray excess in the X-ray
spectrum of the classical high-ionization Seyfert 2 galaxy Tol
0109--383. The X-ray emission of this source observed below 7 keV is
dominated by reflection from both cold and ionized gas, as seen in the
ASCA data. The excess hard X-ray emission is presumably due to the
central source absorbed by an optically thick obscuring torus with \nH
$\sim 2\times 10^{24}$\psqcm. The strong cold X-ray reflection, if it
is produced at the inner surface of the torus, is consistent with the
picture where much of the inner nucleus of Tol 0109--383 is exposed to
direct view, as indicated by optical and infrared properties. However,
the X-ray absorption must occur at small radii in order to hide the
central X-ray source but leave the optical high-ionization emission
line region unobscured. This may also be the case for objects like
the Seyfert 1 galaxy Mrk231.
\end{abstract}


\section{introduction}

It is widely accepted that obscuration is the key for explaining the
Seyfert 2 properties in active galaxies (Antonucci
1993).  However, the obscuring material in Seyfert 2 nuclei might take various
forms or have a complex structure, and its distribution appears to be
spread over a wide range of radii (e.g., sub-pc to a few hundred pc),
as argued by several authors (e.g.,
Malkan et al 1998; Taniguchi \& Murayama 1998; Matt 2000).  Depending 
on the observational
techniques employed, the obscuration inferred of an active nucleus could be
different. In particular, optical/near-infrared results often differ
from X-ray results (e.g., Goodrich, Veilleux \& Hill 1994) as they
probe different phenomena taking place at different optical depths.
This can, in turn, be used to investigate the structure of obscuration
in Seyfert 2 nuclei.

The obscuring matter illuminated by a hidden active nucleus in
Seyfert 2 galaxies is
an important source of emission over a wide range of wavelengths.
While dust reradiation in the far-infrared and some strong optical emission
lines, such as [OIII]$\lambda 5007$, are considered to be
isotropic, any illuminated surface of the obscuring matter
could be viewing-angle dependent if it is in the form of a torus
(Heckman 1995).
High-ionization emission lines (Pier \& Voit 1995; Murayama \& Taniguchi 1998), 
hot ($T_{\rm eff}\sim 1000$ K) dust emission in the mid-infrared
(e.g., Pier \& Krolik 1992; Murayama, Mouri \& Taniguchi 2000) 
and X-ray reflection (Ghisellini, Matt \&
Haardt 1994) are among the observables expected from an illuminated
inner surface of an optically-thick torus. These are therefore potential
indicators of orientation of the obscuring matter in Seyfert 2 nuclei.
Heisler, Lumsden \& Bailey (1997) also argued that the detectability of 
polarized broad-line regions (PBLR) in Seyfert 2 nuclei is related
to our viewing angle of the obscuring torus (see also 
Taniguchi \& Anabuki 1999).

X-ray data such as that obtained from BeppoSAX and ASCA have proved
to be a powerful probe to heavily obscured active nuclei.  Hard
X-ray observations available from BeppoSAX penetrate through a
large optical depth (e.g., Maiolino et al 1998).  X-ray reflection is,
as mentioned above, an important diagnostic of the matter directly
visible to us and its complex spectral features are better
investigated with the reasonable spectral resolution available from
ASCA. Here, we present a combined ASCA/BeppoSAX broad band X-ray
spectrum of the Seyfert 1.8 nucleus of Tol 0109--383, which exhibits
optical properties resembling those of a Seyfert 1 nucleus in
direct view, and yet has a central X-ray source hidden behind an
extremely thick absorber, demonstrating segregation of the nuclear
obscuration.

\section {Tol 0109--383 (=NGC424)}

Tol 0109--383 is a southern early-type spiral galaxy at a redshift of
0.0117 (Da Costa et al 1991). This galaxy hosts one of the brightest
Seyfert 2 nuclei exhibiting a very high ionization optical spectrum,
inferred by the presence of [FeVII]$\lambda 6087$,
[FeX]$\lambda 6374$ and [FeXIV]$\lambda 5303$ (Fosbury \& Sansom 1985;
Durret \& Bergeron 1988; Murayama, Taniguchi \& Iwasawa 1998).  The
high ionization region of Tol 0109--383 traced by [FeX]$\lambda 6374$
has been found to extend up to 5 arcsec (corresponding to 1.1 kpc for
the source distance of 45.3 Mpc\footnote{The distance is derived
assuming $H_0 = 75$ \kmpspMpc and a recession velocity with respect to
the Galactic Standard of Rest, $V_{\rm GSR}=3397 $ \kmps (de
Vaucouleurs et al 1991)}) in radius (Murayama et al 1998), although
most ($\sim 70$ per cent) of the emission is concentrated in the inner
1 arcsec (220 pc) in radius. 

An obscured Seyfert 1 nucleus has been suggested for the nuclear activity
in Tol 0109--383 (e.g., Boisson \& Durret 1986).
Broad Balmer emission has been detected in polarized light 
(Moran et al 2000) as well as in direct light
(Boisson \& Durret 1986; Murayama et al 1998).
Permitted FeII, which is often seen in the spectra of Seyfert 1 galaxies
but rarely seen in Seyfert 2 nuclei, has also been detected 
(Murayama et al 1998).
The reddening to the narrow-line regions (NLR) has been estimated to be 
$A_{\rm V}= $1.4--2.0 (Fosbury \& Sansom 1983; Durret \& Bergeron 1988;
Murayama et al 1998) based on the Balmer decrement, which is comparable to
the mean value ($A_{\rm V}=1.1$) for Seyfert 2 galaxies (De Robertis \& 
Osterbrock 1986). The Hubble Space Telescope / Wide Field Planetary Camera 2 
(HST/WFPC2) image shows clear evidence for a dust lane
running across the central part of the galaxy (Malkan et al 1998).
This could provide an obscuring screen in front of the nuclear region. 


The radio power (e.g., log $P_{20\rm cm}\approx 22$ W Hz$^{-1}$, Ulvestad \&
Wilson 1989),
infrared luminosity ($L_{8-1000\mu m}=1.3\times 10^{44}$ \ergps, or
$3.4\times 10^{10}$\Ls, IRAS FSC) and weak UV luminosity ($L_{1450}=
1.9\times 10^{41}$\ergps, a 100\AA-wide band IUE measurement 
from Mulchaey et al 1994) are typical of Seyfert
2 galaxies. A notable property is the very warm IRAS colour 
(e.g., $S_{60}/S_{25}=1.0$), indicating the presence of hot dust.
Also, the small size of the radio source, which is only slightly resolved
with VLA (at 6 and 20 cm, Ulvestad \& Wilson 1989; at 3.6 cm, 
Threan et al 2000) and ATCA (at 3.5 cm, Morganti et al 1999), 
is not typical of Seyfert 2 galaxies.
Detection of soft X-ray emission 
with the Einstein Observatory (Green, Anderson \& Ward
1992) and ROSAT (Rush et al 1996; Voges et al 1999) has been reported.
However, the origin of the soft X-ray emission has not been clear 
(Murayama et al 1998). Prior to the present paper, Collinge \& Brandt (2000)
analyzed the ASCA data and concluded that Tol 0109--383 is a Compton-thick
source.


\section{Observations and data reductions}


\begin{table*}
\begin{center}
\caption{ASCA and BeppoSAX observations of Tol 0109--383. The count rates
of two same type of detectors, the SIS (S0/S1) and GIS (G2/G3), are shown. }
\begin{tabular}{lclccc}
Satellite & Date &  Instrument & Energy range & Count rate & Exposure \\
&&&& \cps & $10^3$ s \\[5pt]
ASCA & 1997 July 02 & SIS(1CCD/Faint) & 0.6--7 keV & $(2.0/1.5)\times 10^{-2}$ & $47\times 2$ \\
     &              & GIS (PH)        & 0.9--10 keV & $(1.1/1.5)\times 10^{-2}$ & $31\times 2$ \\[5pt]
BeppoSAX & 1999 July 26--28 & LECS & 0.2--4 keV & $2.3\times 10^{-3}$ & 25 \\
         &                  & MECS & 2--10 keV & $8.9\times 10^{-3}$ & 64 \\
         &                  & PDS  & 15--100 keV & 0.16 & 32 \\
\end{tabular}
\end{center}
\end{table*}

Tol 0109--383 was observed with ASCA (Tanaka, Inoue \& Holt 1994) on
1997 July 2, and subsequently with BeppoSAX (Boella et al 1997a) on 1999
July 26--28.  Details of the observations are summarised in Table 1.

The four detectors on board ASCA, the Solid-state Imaging
Spectrometers (SIS; S0 and S1) and the Gas Imaging Spectrometers (GIS;
G2 and G3), were operating normally. The Seyfert galaxy was centred at
the nominal position on the best-calibrated CCD chips of each SIS
detector (S0/C1 and S1/C3). Standard Revision-2 calibration and data
reduction technique were employed, using FTOOLS version 4.2 provided
by the ASCA Guest Observer Facility at NASA's Goddard Space Flight
Center. The Seyfert galaxy is the brightest source in the detector
field of view and there is no other source contaminating the source
extraction region in each detector. Background data were taken from
a source-free region on the same detector in the same observation.
Significant decrease in efficiency of the SIS in the soft X-ray band 
has been reported in ASCA observations carried out after 1994 (ASCA GOF 1999).
The detector responses used here have not been corrected for this effect, but 
for a faint source like Tol 0109--383, the statistical error overwhelms
systematic error in the responses so that the results presented
in this paper is not seriously affected.

The BeppoSAX observation provides a dataset from three detectors; the
Low Energy Concentrator Spectrometer (LECS, Parmar et al 1997), the
Medium Energy Concentrator Spectrometer (MECS, Boella et al 1997b) and
the Phoswitch Detector System (PDS, Frontera et al 1997).  The data
from each detector obtained through standard data reduction provided
by the SAX Data Centre are used. The background data for the LECS and
MECS were taken from the deep blank field observations while the
off-source data during the observation are used as the background for
the PDS which is a collimated instrument rocking between on and
off-source positions.

We quote in this
paper values of X-ray flux obtained from the ASCA GIS and BeppoSAX
MECS, for which the absolute flux calibration is reliable; indeed the
two detectors are in good agreement with each other. The ASCA SIS
agrees with the GIS in spectral shape but shows a few per cent
smaller normalization than the GIS. The standard relative
normalization factor 0.86 is assumed for the PDS. 
The normalization factor found for the LECS is $\sim 0.7$, which is 
within the reasonable range.

\section{X-ray spectrum}


\begin{figure}
\centerline{\psfig{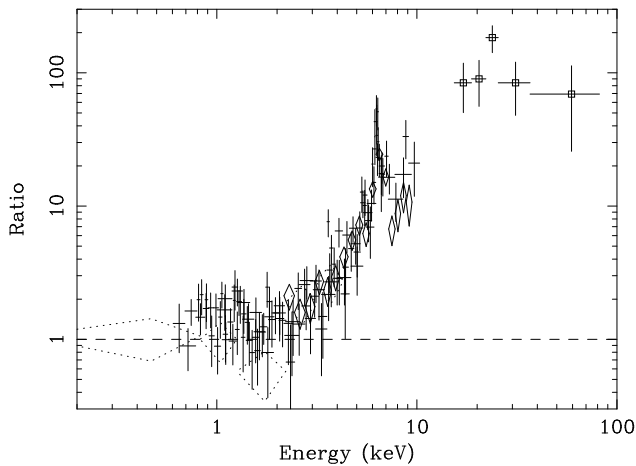}}
\caption{The data from all the detectors of BeppoSAX (diamonds: LECS 
(dotted) and MECS (solid);
and squares: PDS) and ASCA (crosses: SIS and GIS), devided by a power-law of 
$\Gamma = 2.0$ modified by the Galactic absorption (\nH $=1.8\times 10^{20}$
\psqcm). The power-law has been normalized to match the data at 1 keV,
where the luminosity is $4\times 10^{40}$erg s$^{-1}$keV$^{-1}$ for Tol 0109--383.}
\end{figure}

\subsection{Broad-band X-ray spectrum}

Fig. 1 shows the data from all the detectors of BeppoSAX and ASCA,
divided by a power-law with a photon-index of $\Gamma = 2$, to
demonstrate the broad-band spectral structure of Tol 0109--383.  The
photon-index $\Gamma = 2.0$, typical of Seyfert nuclei and QSOs, is
tentatively used here to demonstrate the energy distribution of the
X-ray source (see also similar plots for NGC1068, NGC4945 and NGC6240
in Guainazzi et al 2000 and Vignati et al 1999 for
comparison). Because of the complexity of the X-ray spectrum and the
present quality of the data, the spectral slope of the central source
in this object cannot be constrained very well, but $\Gamma =2$ is
consistent with the data and we use this power-law slope hereafter
(see also Section 4.3). The power-law has been modified by Galactic
absorption (\nH = $1.8\times 10^{20}$\psqcm, Dickey \& Lockman 1990)
and normalized to match the observed flux at 1 keV.  The luminosity at
1 keV is about $4\times 10^{40}$erg s$^{-1}$keV$^{-1}$ for our assumed
source distance of 45.3 Mpc.

The continuum spectrum becomes hard above 3 keV: the 3--6 keV
continuum has a photon-index of $\Gamma = -0.28^{+0.44}_{-0.48}$,
obtained by fitting jointly the ASCA four detectors and the MECS with
a power-law.  A strong iron K emission line is evident at $\sim 6.4$
keV (although the BeppoSAX data show a higher line energy, we believe
that the line is at 6.4 keV, as found in the ASCA data, which is
discussed in detail in Section 4.3). These are consistent with a
reflection spectrum from cold matter.

However, cold reflection alone cannot explain the whole range of
observed broad-band spectrum.  The cold reflection model, e.g., {\tt
pexrav}, Magdziarz \& Zdziarski (1995), with an illuminating power-law
source of $\Gamma = 2$, to match the Fe K band data leaves significant
surplus emission above 7 keV observed with the GIS and the PDS. This hard
X-ray excess is not prominent in the SIS data due to the sharply declining
detector response at those energies. The
large hard X-ray excess suggests the presence of a
strongly absorbed component with a column density over
$10^{24}$\psqcm, similar to Circinus Galaxy (Matt et al 1999), NGC6240
(Vignati et al 1999) and NGC4945 (Iwasawa et al 1993; Done et al 1996;
Guainazzi et al 2000; Madejski et al 2000).  In the soft X-ray band,
marginal evidence for emission-line features are found below 3 keV in
the ASCA SIS spectrum, which require reflection from partially ionized
gas.  Details of the spectral components mentioned above are given in
the following sections.



\subsection{The ASCA data}


\begin{figure}
\centerline{\psfig{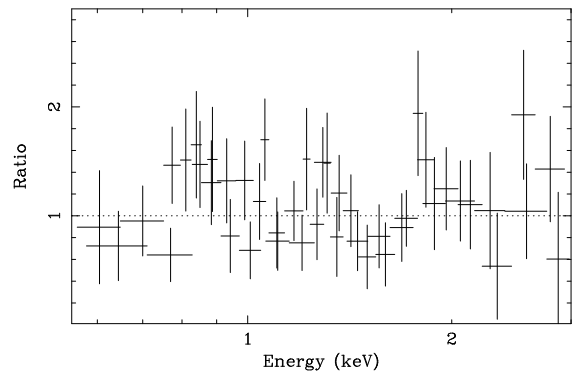}}
\caption{The ratio of the SIS data and a power-law model with $\Gamma =2$.
Possible emission features are seen at 0.87, 1.32, and 1.86 keV.}
\end{figure}

As Fig. 1 shows, the 0.6--3 keV spectrum can be described roughly with
a power-law of $\Gamma\sim 2$. Although a simple power-law fit
provides a reasonably good fit ($\chi^2 = 64.0$ for 68 degrees of
freedom with $\Gamma = 2.1\pm 0.3$) to the SIS data, possible
line-like features can also be seen (Fig. 2).  Only a feature at
$0.87\pm 0.04$ keV is detected with over 90 per cent significance
according to the F-test when a narrow gaussian is fitted, while the
other two features at $1.32\pm 0.06$ and $1.86\pm 0.16$ keV are less
significant.  The equivalent widths of these line features range
between 80--140 eV.  The 0.87 keV feature can be due to a OVIII
recombination continuum (see Griffiths et al 1998 for a similar
feature in the ASCA spectrum of Mrk 3 where they inferred $\xi\sim
500$ for the reflecting matter) or part of the Fe-L blend (e.g.,
FeXVII). The features at 1.32 and 1.86 keV are consistent with
recombination lines from MgXI(1.34 keV) and SiXIII(1.85 keV),
respectively. These features in the soft X-ray band could also
originate from thermal emission form a starburst. However, the optical
and infrared properties of this object suggest that starburst activity is
weak even if it is persent.

The Fe K line is found at a rest-energy of $6.37\pm 0.03$ keV.
No significant broadening is found (the 90
per cent upper limit is 110 eV in the gaussian dispersion, $\sigma$).
The line flux is $2.1^{+0.5}_{-0.4}\times 10^{-5}$\phpspsqcm, and 
the corresponding equivalent width is $\simeq 1.1$ keV with respect to the
neighbouring flat continuum.

As mentioned in the previous section, the ASCA spectrum can be described 
with a sum of reflections from cold and ionized matter (also see Matt et al
2000). The Fe K band spectrum is well explained by reflection from thick,
cold matter. The hard continuum fits well the reflection spectrum of 
{\tt pexrav}. 
The soft X-ray emission-line features could originate
from photoionized gas which also reflects a small fraction of the continuum 
light of the hidden central source. In this case, the photoionized gas 
should be optically thin, and the ionization parameter $\xi$ is 
a few hundreds.

Reflection from mildly ionized, optically thick matter is a
possible alternative to explain the whole ASCA band spectrum, as proposed for
the Circinus Galaxy (Bianchi et al 2001). 
Note that the Fe K line remains at 6.4 keV and some excess soft X-ray
emission is produced when $\xi\leq 30$. 
However, this possibility does not appear to be the case in Tol 0109--383.
We have compared the $\xi=30$ reflection spectrum of 
Ross, Fabian \& Young (1999) with the data. The data above 3 keV
are too hard to match the model spectrum, indicating that the Fe K band 
emission is produced in matter significantly less ionized
than $\xi=30$. Therefore a combination of reflection both 
from cold and highly ionized gas is more likely. 

\subsection{The BeppoSAX data}

\begin{figure}
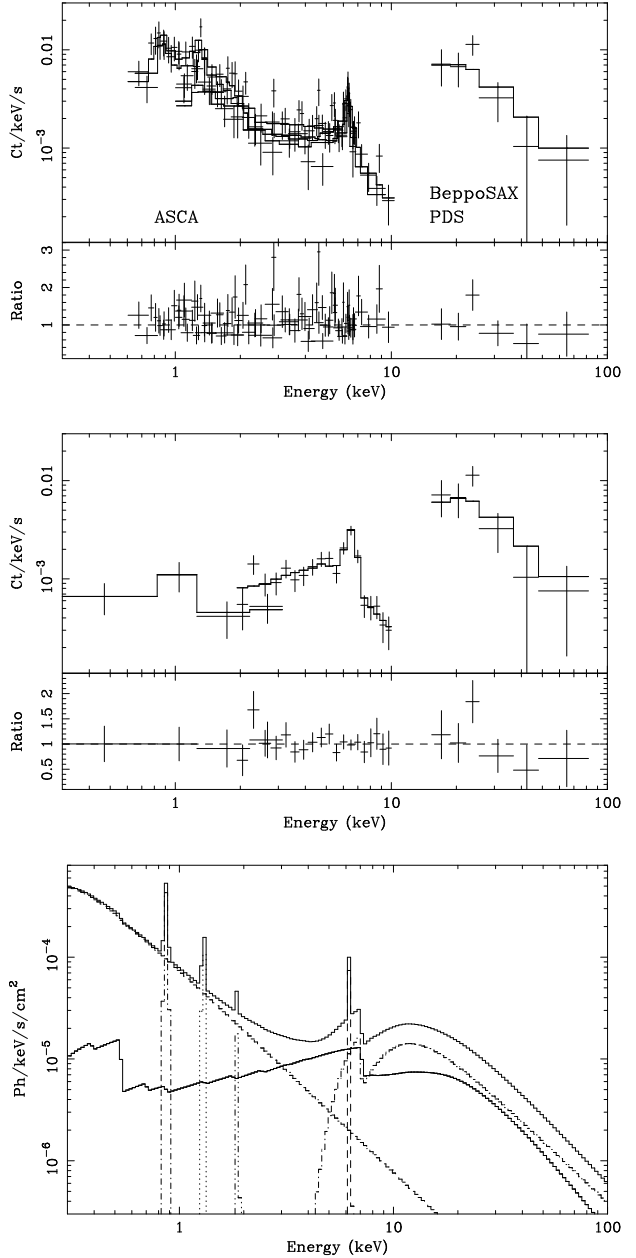

\centerline{\psfig{figure=fig3a.ps,width=0.46\textwidth,angle=270}}
\vspace{4mm}
\centerline{\psfig{figure=fig3b.ps,width=0.46\textwidth,angle=270}}
\vspace{4mm}
\centerline{\psfig{figure=fig3c.ps,width=0.46\textwidth,angle=270}}
\caption{Upper panel: the ASCA data from the four detectors and the
BeppoSAX PDS data of Tol 0109--383, fitted by the model consisting of
cold and warm reflection and a strongly absorbed power-law. Gaussian
lines at the rest-energies of 0.87, 1.34, 1.86 and 6.4 keV are also
included. Middle panel: The BeppoSAX LECS, MECS and PDS data of Tol
0109--383, fitted with the same model as above apart from the
absorption column density and iron K line parameters (see text). Lower
panel: The best-fit model for the ASCA+PDS data, consisting of
components given Table 2.}
\end{figure}

\begin{table*}
\begin{center}
\caption{The spectral fits to the ASCA and BeppoSAX data of Tol 0109--383. In fitting, free parameters are normalizations of respective
components, the absorption column density, and the Fe K line parameters. The line energies of the three soft X-ray lines are fixed.
While the line fluxes are left as free parameters in the fit of the ASCA data, they are fized at the ASCA values for the fit to the BeppoSAX
data. All the spectral components are modeified by Galactic absorption of \nH = $1.8\times 10^{20}$\psqcm. The line width of the Fe K
line is measured by gaussian dispersion in keV.}
\begin{tabular}{lll}
Spectral components & Model & \\[5pt]
Soft X-ray emission & \multicolumn{2}{l}{Power-law ($\Gamma = 2.0$)} \\
 & \multicolumn{2}{l}{3 narrow gaussian lines (at 0.87, 1.36 and 1.86 keV)} \\
Cold reflection & \multicolumn{2}{l}{{\tt PEXRAV} ($\Gamma = 2.0$, reflection only)}  \\
 & \multicolumn{2}{l}{A gaussian line (Fe K); see the separate table} \\
Absorbed central source & \multicolumn{2}{l}{Power-law ($\Gamma = 2.0$)} \\
 & Cold absorption: & \nH = $1.6^{+1.0}_{-1.0}\times 10^{24}$\psqcm\thinspace 
(ASCA+PDS)\\ 
& & \nH = $3.5^{+1.8}_{-1.4}\times 10^{24}$\psqcm\thinspace (SAX) \\
\end{tabular}
\vspace{5mm}
\begin{tabular}{lccc}
Fe K line & Energy & Line width & Line flux \\
Measurements & (keV) & (keV) & \phpspsqcm \\[5pt]
ASCA & $6.37\pm 0.03$ & $< 0.11$ & $2.1^{+0.5}_{-0.4}\times 10^{-5}$ \\
BeppoSAX & $6.63\pm 0.09$ & $0.22^{+0.11}_{-0.12}$ & $2.6^{+1.0}_{-0.8}\times 10^{-5}$ \\
\end{tabular}
\begin{tabular}{ccc}
Quality of fit & $\chi^2$ / dof \\[5pt]
ASCA+PDS & 187.3 / 186 \\ 
BeppoSAX & 16.1 / 18 \\[5pt]
\end{tabular}
\end{center}
\end{table*}

\begin{table*}
\begin{center}
\caption{High excitation lines and infrared colours of Compton-thick Seyfert 2
galaxies. Optical Seyfert type is given in the second column. 
PBLR represents a Seyfert 2 galaxy with Polarized Broad-Line Region.
The values of [FeVII]$\lambda 6087$/[OIII]$\lambda 5007$ have not
been corrected for internal reddening. $S_{60}/S_{25}$ and $S_{25}/S_{12}$
are the ratio of 
IRAS flux densities at 60 and 25$\mu$m, and 25 and 12$\mu $m.
$R(L,25)$ is the mid-infrared diagnostic of a dusty torus 
introduced by Murayama
et al (2000), as defined by $R(L,25)=log[(\nu_{3.5\mu m}S_{\nu 3.5})/
(\nu_{25\mu m})S_{\nu 25})]$. 
References::$<$[FeVII]/[OIII]$>$ 1: Murayama et al 1998; 2: Nagao et al 2000;
3: Alloin et al 1992; 4: Kleinmann et al 1988; 5: Tran et al 2000; $<$IRAS flux$>$ 6: IRAS Faint Source Catalogue; 7: IRAS Team 1983; $<$L-band photometry$>$ 4: Kleinmann et al 1988; 8: Glass \& Moorwood 1985; 9: Murayama et al 2000.
$\dag$ Contamination from the host galaxy is likely. $\ddag$ The $L$ band photometry is approximated by the $L'$ data.}
\begin{tabular}{lccccl}
Galaxy name & Class & [FeVII]/[OIII] & $S_{60}/S_{25}$ & 
$R(L,25)$ & Ref.\\[5pt]
Tol 0109--383 & PBLR & $5.36\times 10^{-2}$ & 1.03 & $-0.036$ & 1,6,8 \\ 
NGC 1068 & PBLR & $1.87\times 10^{-2}$ & 2.07 & $-0.79$ & 2,6,9 \\
NGC 7674 & PBLR & $1.59\times 10^{-2}$ & 2.94 & $-0.76^{\dag}$ & 2,6,9\\
IRAS 09104+4109 & PBLR & $1.95\times 10^{-2}$ & 1.57 & $-1.25^{\ddag}$ & 4,5,6\\
Circinus galaxy & S2 & $1.69\times 10^{-2}$ & 4.31 & $-1.16^{\dag}$ & 2,7,9\\
NGC 5643 & S2 & --- & 5.34 & $-1.26$ & 2,6,8\\
ESO 138-G1 & S2 & $2.71\times 10^{-2}$ & --- & --- & 3\\
Mrk 231 & S1 & --- & 3.69 & $-0.53$ & 6,9 \\
\end{tabular}
\end{center}
\end{table*}

The data obtained from the LECS and MECS are in good agreement with 
the ASCA data below the Fe K band.
The MECS data show 
a higher centroid-energy of $6.63\pm 0.09$ keV with significant
broadening ($\sigma = 0.22^{+0.11}_{-0.12}$ keV) for the Fe K line. 
A slightly higher line intensity of 
$2.6^{+1.0}_{-0.8}\times 10^{-5}$\phpspsqcm\ is also obtained.
A likely cause of the line broadening would be a blend of a 6.4 keV line
and a higher ionization line, e.g., FeXXV (6.7 keV), which should
be resolved with the ASCA SIS, owing to the higher spectral resolution 
than the MECS and the GIS. However,
no evidence for a higher energy lines is found in the ASCA data.
This could be dues to the gain shift problem with the MECS, although
the offset is larger than that usually found (see e.g., Dupke \&
Arnaud 2001).  Another possible reason for this is the spectral
resolution which is not suitable for a complicated spectrum with a
sharp drop on the higher energy side of the line due to the Fe K
absorption edge. This could shift the line centroid to higher energy
in a low resolution detector. A spectral analysis of the data from the
ASCA GIS only, which is a similar detector to the MECS, gives the line
energy of $6.44^{+0.37}_{-0.10}$ keV.  The large error for the GIS
line energy towards higher energies is suggestive but not conclusive
enough to settle the problem.  Since a narrow line at 6.4 keV is more
plausible on the physical ground given the continuum shape, and an
unexpected gain drift in the detector could lead to the line
broadening, we are cautious about an immediate interpretation of the
discrepancy.

The excess hard X-ray emission seen in the PDS data is most probably due to 
a strongly absorbed primary source.
The 7--10 keV MECS flux is marginally lower (by $\sim 30$ per cent) 
than that measured by ASCA,
which could be attributed to an intrinsic change in 
the low energy tail of the absorbed component. Spectral fits to 
the ASCA data (with the BeppoSAX PDS data) and the BeppoSAX
data are presented in Fig. 3. Apart from the Fe K line and absorption
column density, the fitted models are identical. 
The unabsorbed source is assumed to have 
a power-law spectrum with $\Gamma = 2$ and
no high energy cut-off.
We utilized {\tt pexrav} and a gaussian Fe K line 
to model the cold reflection. The reflecting slab is assumed to
be of Solar abundance and nearly face-on ($\sim 20^{\circ}$ in inclination).
The soft X-ray continuum is described by a $\Gamma = 2$ power-law
modified only by Galactic absorption.
The X-ray absorption column density to the central source is estimated 
to be $(1.6\pm 1.0)\times 10^{24}$\psqcm\ from the ASCA+PDS and
$3.5^{+1.8}_{-1.4}\times 10^{24}$\psqcm\ from the three detectors of
BeppoSAX.  Since the ASCA and BeppoSAX observations are not simultaneous, 
the slightly smaller \nH\ value obtained from the ASCA+PDS fit might
suggest a luminosity increase of the central source, assuming that the
column density is not variable.
The three continuum components plus the gaussian lines described above (and in Table 2)
fit the data well ($\chi^2=187.3$ for 186 degrees of freedom for ASCA+PDS;
$\chi^2=16.1$ for 18 degrees of freedom for LECS+MECS+PDS), as
shown in Fig. 3.  
Although the data do not provide a good constraint on the continuum slope
if it is left as a free parameter,
a flat continuum with a photon-index smaller than 1.5 can be ruled out.
The effect of Compton scattering in this large absorbing column 
(e.g., Matt, Pompilio \& La Franca 1999) is 
not taken into account here. An appropriate correction will be applied on
estimating the absorption-corrected luminosity.

The observed 0.5--2 keV, 2--10 keV, and 20--100 keV fluxes are 
$2.2\times 10^{-13}$\ergpspsqcm, $1.6\times 10^{-12}$\ergpspsqcm
and $1.8\times 10^{-11}$\ergpspsqcm, respectively.

\section{discussion}

The ASCA spectrum is consistent with a reflection-dominated emission,
as concluded by Collinge \& Brandt (2000).
The BeppoSAX PDS detection of a transmitted component at higher energies
allows us to measure the column density of the Thomson-thick
absorber 
(\nH $\sim $2--3$\times 10^{24}$\psqcm, or Thomson depth $\tau_{\rm T}\sim 2$,
assuming Solar abundance). 
The best estimate of the absorption-corrected 2--10 keV 
luminosity of the central source is 
$7\times 10^{42}\gamma $\ergps\ for our assumed source distance.
With the flattest limit of the spectral slope of the central source,
$\Gamma = 1.5$, this value is a factor of 2 larger.
Note that the correction for Compton-scattering (Matt et al 1999)
has been made ($\gamma = 1$ for a spherical obscuration; the value increases to
$\sim 3$ if the covering fraction is small for \nH =$3\times 10^{24}$\psqcm). 

The reflected fraction (i.e., the luminosity ratio of 
the observed and the absorption-corrected 
continuum) in the 2--10 keV band,
where most of the observed flux comes from cold reflection, is
$\sim 5\gamma^{-1}$ per cent. Compared with the albedo from cold matter,
this means that a quarter to half of the reflection
expected from a slab subtending at $2\pi$ is visible to us.
If the cold reflection occurs at the inner surface of the obscuring
torus, then a large fraction of the surface is exposed to our view.

The optical polarization of the nucleus of Tol 0109--383 is relatively
high (1.5 per cent, Brindle et al 1990; 2.3 per cent, Moran et al 2000 in the
V band) and polarized broad emission lines have been detected (Moran et al 
2000). 
Note that broad Balmer emission has been detected even in the direct
light in Tol 0109--383 (Murayama et al 1998).
Several authors have argued that a moderately inclined obscuring torus is
prefered for detecting PBLR (e.g., Heisler et al 1997; 
Taniguchi \& Anabuki 1999). 
The PBLR Seyfert 2 sample of Heisler et al (1997)
tends to have warm IRAS colours, which is naturally explained by
the viewing angle effect, i.e., hot dust at inner radii is more visible
in a more inclined system (e.g., Pier \& Krolik 1992). 
We have verified that this trend holds for the 
distance-limited spectropolarimetric survey sample of Moran et al (2000).
Awaki et al (2000), however, prefer a large size of the scattering 
regions to the viewing angle effect for detectability of the PBLR.

For Tol 0109--383, the viewing angle appears to be an important key.
Observed properties listed below all point to us looking into the
inner part of the active nucleus: 1) the relatively large X-ray
reflection fraction from cold matter, as discussed above; 2) the
detection of broad Balmer emission and permitted FeII emission in
direct light (Murayama et al 1998); 3) the very warm IRAS colour (see
Table 3); and 4) the presence of high-ionization emission lines,
particularly the high [FeVII]$\lambda 6078$/[OIII]$\lambda 5007$ ratio
(see Table 3).  

As mentioned in the Introduction, some high ionization
emission lines may originate in the inner surface of the optically
thick torus (e.g., Pier \& Voit 1995; Murayama \& Taniguchi 1998),
thus could be an indicator of the torus viewing angle. Nagao et
al (2000) have shown that the [FeVII]/[OIII] ratio is markedly
different between the two types of Seyferts. 
A good segregation is also seen with the mid-infrared flux ratio
of 3.5$\mu $m ($L$ band) and 25$\mu$m (Murayama, Mouri \& Taniguchi 2000).
The IRAS $S_{60}/S_{25}$ ratio may not necessarily be a good indicator
of torus orientation (Fadda et al 1998; Murayama, Mouri \& Taniguchi
2000; Alexander 2001) since there must be a significant contribution
to the $60\mu $m flux from the host galaxy, although Heisler et al
(1997) show that the PBLR and non-PBLR Seyfert-2 galaxies are well
separated using this ratio. Based on the compact dusty torus model by
Pier \& Krolik (1992), Murayama et al (2000) demonstrate that the flux
ratio of L-band and 25$\mu $m is a good diagnostic to the torus
orientation, provided contamination from starburst (or a host galaxy)
is negligible.  

We show the values of possible orientation measures
mentioned above ([FeVII]/[OIII] ratio, $S_{60}/S_{25}$ and
$R(L,25)=log[(\nu_{3.5\mu m}S_{\nu 3.5})/(\nu_{25\mu m})S_{\nu 25})]$)
of Tol 0109--383 along with other Compton-thick Seyfert galaxies
(Collinge \& Brandt 2000; Matt et al 1997; Matt et al 1999; Risaliti
et al 1999; Malaguti et al 1999; Iwasawa et al 2000; Maloney \&
Reynolds 2000), some of which have been known as PBLR Seyfert 2s, in
Table 3. Two other Compton-thick sources, NGC4945 and NGC6240, are not
included, since a starburst dominates the optical to mid-infrared wave
bands in those galaxies. The high [FeVII]/[OIII] ratio and the flat
infrared spectrum, implied from the small $S_{60}/S_{25}$ and large
$R((L,25)$ of Tol 0109--383 are consistent with a picture that the
interior of the torus is largely exposed, compared to the other
Compton-thick Seyfert 2s.  The lack of correlation between the X-ray
absorption column density (Bassani et al 1999) and the [FeVII]/[OIII]
ratio (Nagao et al 2000) probably means that the structure of
obscuring torii in Seyfert 2 galaxies is not unique.

Despite the close resemblance to a Seyfert 1 nucleus, the central
X-ray source in Tol 0109--383 is hidden behind heavy obscuration.  The
large column of absorbing gas is unlikely to be associated with the
dust lanes imaged by the HST (Malkan et al 1998). This is clearly a
geometrical effect and the X-ray absorption is likely to occur on
small radii, perhaps on the parsec scale, as originally proposed by
Krolik \& Begelman (1988) and also by Matt (2000). More exposure of an
optical Seyfert 1 nucleus is found in Mrk 231, for which an X-ray
nucleus hidden by a Compton-thick absorber has been suggested by
Maloney \& Reynolds (2000). The location of very thick X-ray 
absorbing matter may be distinct from the torus hiding the BLR 
(or an optical nucleus in Seyfert 2 galaxies), and could be in a
special form (e.g., Elvis 2000).

\section*{Acknowledgements}

We thank the BeppoSAX and ASCA teams for their efforts on operation of
the satellites, and the calibration and maintenance of the software.
ACF and KI thank the Royal Society and PPARC for support,
respectively.


\begin{thebibliography}{}

\bibitem{} Alexander D.M., 2001, MNRAS, 320, L15
\bibitem{} Alloin D., Bica, E., Bonatto C., Prugniel P., 1992, A\&A, 266, 117
\bibitem{} Antonucci R.R.J., 1993, ARAA, 31, 473
\bibitem{} \hbox{ASCA Guest Observer Facility (GOF), 1999, http://heasarc.gsfc.}nasa.gov/docs/asca/watchout.html
\bibitem{} Awaki H., Ueno S., Taniguchi Y., Weaver K.A., 2000, ApJ, 542, 175
\bibitem{} Bassani L., Dadina M., Maiolino R., Salvati M., Risaliti G., Della Cecca R., Matt G., Zamorani G., 1999, ApJS., 121, 473
\bibitem{} Bianchi S., Matt G., Iwasawa K., 2001, MNRAS, 322, 669
\bibitem{} Boella G., Butler R.C., Perola G.C., Piro L., Scarsi L., Bleeker J.A.M., 1997a, A\&AS, 122, 299
\bibitem{} Boella G., et al, 1997b, A\&AS, 122, 327
\bibitem{} Boisson C., Durret F., 1986, A\&A, 168, 32
\bibitem{} Brindle C., Hough J.H., Bailey J.A., Axon D.J., Ward M.J., Sparks W.B., McLean I.S., 1990, MNRAS, 244, 577 
\bibitem{} Collinge M.J., Brandt W.N., 2000, MNRAS, 317, L35 
\bibitem{} Da Costa L.N., Pellegrini P.S., Davis M., Meiksin A., Sargent W.L.W., Tonry J.L., 1991, ApJS, 75, 935
\bibitem{} De Robertis M.M., Osterbrock D.E., 1986, ApJ, 301, 727
\bibitem{} de Vaucoleurs G., et al, 1991, Third Reference Catalogue of Bright Galaxies (New York: Springer)
\bibitem{} Dickey J.M., Lockman F.J., 1990, ARAA, 28, 215
\bibitem{} Done C., Madejski G.M., Smith D.A., 1996, ApJ, 463, L63
\bibitem{} Durret F., Bergeron J., 1988, A\&A, 173, 219
\bibitem{} Elvis M., 2000, ApJ, 545, 63
\bibitem{} Fadda D., Giuricin G., Granato G., Vecchies D., 1998, ApJ, 96, 117
\bibitem{} Fosbury R.A.E., Sansom A.E., 1985, MNRAS, 204, 1231
\bibitem{} Frontera F., Costa R., dal Fiume D., Feroci M., Nicastro L., Orlandini M., Palazzi E., Zavattini G., 1997, Proc. SPIE, 3114, 206
\bibitem{} Ghisellini G., Matt G., Haardt F., 1994, MNRAS, 267, 743
\bibitem{} Glass I.S., Moorwood A.F.M., 1985, MNRAS, 214, 429
\bibitem{} Goodrich R.W., Veilleux S., Hill G.J., 1994, ApJ, 422, 521
\bibitem{} Green P.J., Anderson S.F., Ward M.J., 1992, MNRAS, 254, 30
\bibitem{} Griffiths R.G., Warwick R.S., Georgantopoulos I., Done C., Smith D.A., 1998, MNRAS, 298, 1159
\bibitem{} Guainazzi M., Matt G., Brandt W.N., Antonelli L.A., Barr P., Bassani L., 2000, A\&A, 356, 463
\bibitem{} Heckman T.M., 1995, ApJ, 446, 101
\bibitem{} Heisler C.A., Lumsden S.L., Bailey J.A., 1997, Nat, 385, 700
\bibitem{} IRAS Team, 1983, Nat, 303, 480
\bibitem{} Iwasawa K., Koyama K., Awaki H., Kunieda H., Makishima K., Tsuru T., Ohashi T., Nakai N., 1993, ApJ, 409, 155
\bibitem{} Iwasawa K., Fabian A.C., Ettori S., 2000, MNRAS, 321, L15
\bibitem{} Kleinmann S.G., Hamilton D., Keel W.C., Wynn-Williams C.G., Eaeles S.A., Becklin E.E., Kuntz K.D., 1988, ApJ, 328, 161
\bibitem{} Krolik J.H., Begelman M.C., 1988, ApJ, 329, 702
\bibitem{} Madejski G., Zycki P., Done C., Valinia A., Blanco P., Rothschild R., Turek B., 2000, ApJ, 535, L87
\bibitem{} Magdziarz P., Zdziarski A., 1995, MNRAS, 273, 837
\bibitem{} Maiolino R., Salvati M., Bassani L., Dadina M., Della Cecca R., Matt G., Risaliti G., Zamorani G., 1998, A\&A, 338, 781
\bibitem{} Malaguti G., et al, 1998, A\&A, 331, 519
\bibitem{} Malkan M.A., Gorjian V., Tam R., 1998, ApJS, 117, 25
\bibitem{} Maloney P.R., Reynolds C.S., 2000, ApJ, 545, L23 
\bibitem{} Matt G., et al, 1997, A\&A, 325, L13
\bibitem{} Matt G., Pompilio F., La Franca F., 1999, New Astronomy, 4, 191
\bibitem{} Matt G., et al, 1999, A\&A, 341, L39
\bibitem{} Matt G., Fabian A.C., Guainazzi M., Iwasawa K., Bassani L., Malaguti G., 2000, MNRAS, 318, 173
\bibitem{} Matt G., 2000, A\&A, 355, L31
\bibitem{} Moran E.C., Barth A.J., Kay L.E., Filippenko A.V., 2000, ApJ, 540, L73
\bibitem{} Morganti R., Tsvetanov Z.I., Gallimore J., Allen M.G., 1999, A\&AS, 137, 457
\bibitem{} Mulchaey J.S., Koratkar A., Ward M.J., Wilson A.S., Whittle M., Antonuncci R.R.J., Kinney A.L., Hurt T., 1994, ApJ, 436, 586
\bibitem{} Murayama T., Taniguchi Y., Iwasawa K., 1998, AJ, 115, 460
\bibitem{} Murayama T., Taniguchi Y., 1998, ApJ, 497, L9
\bibitem{} Murayama T., Mouri H., Taniguchi Y., 2000, ApJ, 528, 179
\bibitem{} Nagao T., Taniguchi Y., Murayama T., 2000, AJ, 119, 2605
\bibitem{} Parmar A.N., 1997, A\&AS, 122, 309
\bibitem{} Pier E., Krolik J., 1992, ApJ, 401, 99
\bibitem{} Pier E.A., Voit M.G., 1995, ApJ, 450, 628
\bibitem{} Risaliti G., Maiolino R., Salvati M., 1999, ApJ, 522, 157
\bibitem{} Ross R.R., Fabian A.C., Young A.J., 1999, MNRAS, 306, 461
\bibitem{} Rush B., Malkan M.A., Fink H.H., Voges W., 1996, ApJ, 471, 190
\bibitem{} Tanaka Y., Inoue H., Holt S., 1994, PASJ, 46, L37
\bibitem{} Taniguchi Y., Anabuki N., 1999, ApJ, 521, L103
\bibitem{} Taniguchi Y., Murayama T., 1998, ApJ, 501, L25
\bibitem{} Threan A., Pedlar A., Kukula M.J., Baum S.A., O'Dea C.P., 2000, MNRAS, 314, 573
\bibitem{} Tran H.D., Cohen M.H., Villar-Martin M., 2000, AJ, 120, 562
\bibitem{} Ulvestad J.S., Wilson A.S., 1989, ApJ, 343, 659
\bibitem{} Vignati P., et al, 1999, A\&A, 349, L57
\bibitem{} Voges W., et al, 1999, A\&A, 349, 389
















\end{thebibliography}
\end{document}